\documentstyle[preprint,eqsecnum,aps,epsf]{revtex}

\begin{document}
\draft

\title{Antisymmetry in Strangeness $-1$ and $-2$ Three-Baryon Systems
}

\author{W.~Gl\" ockle} 

\address{Institut f\"ur theoretische Physik II, Ruhr-Universit\"at
 Bochum, D-4478 Bochum, Germany
}
\author{K.~Miyagawa}

\address{Applied Physics, Okayama University of Science,
 1-1 Ridai-cho, Okayama 700, Japan
}
    
\date{\today}
\maketitle

\begin{abstract}
Using the generalized Pauli principle by adding particle labels
to the usual 
space and spin labels a symmetric Hamiltonian and a corresponding
antisymmetric wavefunction is constructed for systems of three baryons
in the strangeness sectors $S=-1$ and $-2$. Applications are the
$\Xi NN-\Lambda\Lambda N$ and $NN\Lambda -NN\Sigma$ systems.
 Minimal sets of generalized  coupled Faddeev equations for
breakup  and rearrangement operators as well as (possible) bound states
are derived which have
the ordinary Pauli principle among identical particles built in.
The equations found confirm our previous sets of coupled Faddeev equations
 for those systems whose derivation was carried through for
 distinguishable particles
and not using the generalized Pauli principle.

\end{abstract}

\pacs{
}

\narrowtext

\abovedisplayskip 7mm
\belowdisplayskip 7mm
\abovedisplayshortskip 7mm
\belowdisplayshortskip 7mm
\jot 5mm   
\newfont{\myfont}{cmti12 scaled \magstep1}

\section{Introduction}

In the three-baryon system $\Xi NN$ a strong transition occurs
into $YYN$  where $Y$ is either a $\Lambda$ or a $\Sigma$. The nucleon
present
in $YYN$ can be either one contained in $\Xi NN$. Clearly the $\Xi NN$
part of the total wavefunction has to be antisymmetrical under exchange
of the two nucleons. So the question arises how is that reflected
in the second part of the wavefunction belonging to $YYN$. Similarly
the second part has to be antisymmetrical under exchange of two
identical hyperons and one has to ask how is that related to the
first part of the wavefunction. There is also the possibility that
$YYN$ appears as $\Lambda\Sigma N$. Those questions (with the
neglection of the
 $\Sigma$-admixture) have been answered in \cite{ours} starting from
a second
quantisation picture of the system. The final aim in \cite{ours} was
to write down
the set of coupled Faddeev equations with the smallest number of
independent Faddeev amplitudes. The argumentation turned out to be
 rather involved and we would like to present a conceptually and  
algebraically simpler
derivation of the same set of coupled equations. Since it is simpler
we will now also include the possibility of a $\Sigma$-admixture.

The $\Xi d$  system has recently received attention \cite{afnan} in that
it might be
a source of information on the $\Lambda\Lambda$ scattering length after
 it has switched
to the unbound  $\Lambda\Lambda N$   system. In that evaluation of those
two coupled 
three-body systems the AGS equations\cite{ags} for  distinguishable
particles have been used
and the free initial and final states have been antisymmetrised
separately.
Since "clean" experimental information on hyperon-hyperon forces
are still very rare due to the lack of hyperon beams and targets
the proposals to extract information on $YN$ and $YY$ forces from
final state interactions have great importance and therefore a clearly
justified
formalism is in order.

Also under the present circumstances  of having no direct access to
$YN$ scattering light hypernuclei
play an equally important role, since they can be solved rigorously
for baryon-baryon forces in all their complexities and thus provide
 an important test for those forces. First steps
have been undertaken in \cite{hypt} for the hypertriton 
and in \cite{hiyama}
for $^4_\Lambda$He and $^4_\Lambda$H.

With respect to investigations of final state interactions among
 hyperons and nucleons the photon induced $K^+ Y$ production on the 
deuteron has been
evaluated recently\cite{yamamu} again using modern $YN$ forces which
correctly
 bind the hypertriton. The  corresponding electron induced
process on the deuteron has been measured recently\cite{reinhold}
and a first calculation appeared\cite{lee}. 
 Certainly
 more work, both experimentally and
 theoretically, will
come up in the near future.

In section II we describe the formalism of antisymmetrisation for
 particles of different types and apply it to the relatively
 simple two-baryon system. The
more complex situation for three baryons with strangeness $-1$ and
$-2$
and the derivation of coupled Faddeev equations
 is worked out in section III. An application to various three-body
breakup processes initiated by a $\Xi$ particle
is described in section IV. Generalised AGS
equations for elastic and rearrangement scattering are described in
section V.  We  summarize in section VI.

\section{Formalism and Application to the Two-Baryon System}

The $\Xi N$ system can strongly convert into two identical baryons,
$\Lambda\Lambda$ or $\Sigma\Sigma$. Clearly the parts of the
wavefunction
describing $\Lambda\Lambda$ or $\Sigma\Sigma$ are antisymmetrical under
exchange of the two
particles, whereas for the part of the wavefunction describing $\Xi N$
there is no symmetry requirement.
Another possible appearance of that strangeness $-2$ system is
$\Lambda\Sigma$
with distinguishable particles. In \cite{ours}  we used the formalism of
 second quantisation
to derive three coupled equations for the coupled $\Xi N$ and
$\Lambda\Lambda$ systems, neglecting a possible $\Sigma$-admixture.
One of the equations
can be dropped at the expense of working with a non-symmetrical
$2\times 2$ potential matrix.

In the $np$ system is has been known since the early days of nuclear
physics
that adding an isospin label to the particle or in other words extending
the space of states by introducing isospin one can require
that the wavefunction for the distinguishable particles $n$ and $p$ has
to be antisymmetrical. The simple mechanism is that in isospin space
there
is a symmetrical and antisymmetrical basis which can be combined with
the corresponding antisymmetrical and symmetrical wavefunction in 
ordinary
and spin space. Therefore no restriction has been introduced by 
requiring
antisymmetry in the enlarged space. Connected to that antisymmetric 
wavefunction
a symmetric Hamiltonian can be written down. 
This idea has been generalized also
long time ago to SU(3) including strangeness on top of isospin.If the
Hamiltonian severely breaks SU(3) symmetry due to significantly
 different masses
and symmetry breaking force parameters  the use of irreducible
 representations of SU(3)
appears not to be of much help and a formally simpler approach can be
used.
Again we think this is known and we display it again for two particles
to lay the ground for the more complex three-baryon system,where we want to
find the way to the smallest set of coupled Faddeev equations.

Let us introduce states  $|\, \alpha\beta >$ with particle labels
$\alpha$ or $\beta$ denoting nucleons or hyperons. Thus
$\alpha=\{N,\Lambda,\Sigma, \Xi\}$.
Note $|\, \alpha\beta >$ stands for $|\, \alpha>_1|\, \beta >_2$
which means that particle 1 is of the type $\alpha$ and 2 is
of type $\beta$. Thus  the transposition $P_{12}$ yields $ P_{12}$
 $|\, \alpha\beta >$
= $|\, \beta\alpha >$. 
 The completeness
 relation in that particle space is obviously

\begin{equation}
\sum_{\alpha \,\beta}
 |\, \alpha \, \beta> <\alpha \, \beta \, |
=1
\label{2.1}
\end{equation}

Since strangeness is a good quantum number the sum is restricted to
run only over physically allowed states. For $S=-1$ these are      
$|\, N\Lambda>$, $|\, \Lambda N>$, $|\, N\Sigma>$, 
 and $|\, \Sigma N>$.
For $S=-2$ one has      
$|\, \Lambda\Lambda>$, $|\, \Sigma\Sigma >$, $|\, \Xi N>$, 
$|\, N\Xi >$, $|\, \Lambda\Sigma >$ and $|\, \Sigma\Lambda>$.
Clearly the  sum in Eq.~(\ref{2.1})is symmetrical under
exchange of
the two particles. Applying that sum onto the corresponding
 two-body Hamiltonian
from both sides
one creates a fully symmetrical Hamiltonian

\begin{equation}
\begin{array}{c}
H \equiv \displaystyle{\sum_{\alpha \beta} \sum_{\alpha ' \beta '}} | 
\alpha \beta >< \alpha \beta | H | \alpha ' \beta ' >< \alpha ' \beta ' |
\\
= \displaystyle{\sum_{\alpha \beta}} | \alpha \beta > ( \frac{p_{1}^{2}}
{2m_{\alpha}} + \frac{p_{2}^{2}}{2m_{\beta}} + m_{\alpha} + m_{\beta} )
 < \alpha \beta | 
\\
+ \displaystyle{\sum_{\alpha \beta} \sum_{\alpha ' \beta '}}
 | \alpha \beta > V_{\alpha \beta , \alpha ' \beta '} 
< \alpha ' \beta ' |
\end{array}
\label{2.2}
\end{equation}

Note that we allow for different masses $m_\alpha$ without
destroying the symmetry of the kinetic energy. In the configuration or
momentum space representation including (possible) spin dependencies
the potentials are denoted by $V_{\alpha \beta , \alpha ' \beta '}(12)$.
In
general these potential functions are not symmetrical. But they have
the property

\begin{equation}
V_{ \beta\alpha ,  \beta '\alpha '}(12)=
V_{\alpha \beta , \alpha ' \beta '}(21)
\end{equation}

\noindent
 Transpositions  applied to potential functions
act on the arguments (momenta and spins) and thus one has

\begin{equation}
P_{12}V_{ \alpha\beta ,  \alpha '\beta '}(12)=
V_{\alpha \beta , \alpha ' \beta '}(21)P_{12}
\end{equation}

\noindent
They do not change  the labels for the potential functions, which are just
parameters of those functions.

To that symmetrical Hamiltonian belongs obviously a fully 
antisymmetrical two-baryon wavefunction:

\begin{equation}
|\Psi>=
\sum_{\alpha \,\beta}
 |\, \alpha \, \beta> <\alpha \, \beta \, |\Psi>
\equiv\sum_{\alpha \,\beta} |\, \alpha \, \beta>
\psi_{\alpha \,\beta} 
\end{equation}

 Like the potential functions the amplitudes $\psi_{\alpha \beta}$ live in spin
 and configuration or
momentum space and corresponding representations will be denoted by
 $\psi_{\alpha \beta}(12)$.  For $\alpha\neq\beta$ one
 introduces
two-particle states

\begin{equation}
|\chi_{\alpha \beta}^{a}> \equiv \frac{1}{\sqrt{2}}
(|\alpha \beta>-|\beta \alpha>)
\label{2.3}
\end{equation}
\begin{equation}
|\chi_{\alpha \beta}^{s} >\equiv \frac{1}{\sqrt{2}}
(|\alpha \beta>+|\beta \alpha>)
\label{2.4}
\end{equation}

\noindent
with obvious symmetry properties. In addition there is the
symmetrical state

\begin{equation}
\chi_{\alpha \alpha}^{s} \equiv |\alpha \alpha>
\label{2.5}
\end{equation}

Then the  total state $\Psi$ can be written as

\begin{equation}
|\Psi>=\sum_{\alpha}
 |\, \alpha \, \alpha> \psi_{\alpha\alpha}
+
\sum_{\alpha\neq\beta}\frac{1}{\sqrt{2}}
(|\chi_{\alpha \beta}^{a}>+|\chi_{\alpha \beta}^{s} >)
 \psi_{\alpha\beta}
\end{equation}

Now we impose antisymmetry

\begin{equation}
P_{12}|\Psi>=-|\Psi>
\end{equation}

\noindent
which  leads after some inspection to

\begin{equation}
P_{12}\psi_{\alpha \beta}=-\psi_{\beta\alpha}
\end{equation}

\noindent
Note $P_{12}$ in Eq. (2.10) acts on the particle labels of all states,
 while
 $P_{12}$ in Eq. (2.11) acts only on those for momentum( configuration)
 and spin
states. In other words Eq. (2.11) leads to

\begin{equation}
\psi_{\alpha \beta}(21)= -\psi_{ \beta\alpha}(12)
\end{equation}

The antisymmetry of (2.9) is explicit if we
keep (2.11) in mind and rewrite (2.9) as

\begin{eqnarray}
|\Psi>&=
\displaystyle\sum_{\alpha} |\, \alpha \, \alpha> \psi_{\alpha\alpha}
&+\sum_{\alpha <\beta} |\chi_{\alpha \beta}^{a}>
\frac{1}{\sqrt{2}}(\psi_{\alpha\beta}- \psi_{\beta\alpha})
\nonumber\\
&
&+\sum_{\alpha <\beta} |\chi_{\alpha \beta}^{s}>
\frac{1}{\sqrt{2}}(\psi_{\alpha\beta}+ \psi_{\beta\alpha})
\nonumber\\
&=
\displaystyle\sum_{\alpha} |\, \alpha \, \alpha> \psi_{\alpha\alpha}
&+\sum_{\alpha <\beta} |\chi_{\alpha \beta}^{a}>
\frac{1}{\sqrt{2}}(1+P_{12})\psi_{\alpha\beta}
\nonumber\\
&
&+\sum_{\alpha <\beta} |\chi_{\alpha \beta}^{s}>
\frac{1}{\sqrt{2}}(1-P_{12})\psi_{\alpha\beta}
\end{eqnarray}

\noindent
Here we introduced a definite but arbitrary ordering of the particle
labels.

Now projecting $H\Psi =E\Psi$ onto $<12|$   $<\alpha \, \beta \,|$
where  $<12|$ stands for momenta and spin states, one obtains

\begin{eqnarray}
&&(\frac{p_{1}^{2}}{2m_{\alpha}} + \frac{p_{2}^{2}}{2m_{\beta}} + m_{\alpha}
 + m_{\beta})
\psi_{\alpha \beta}(12)+\sum_{\alpha' \neq \beta'}
V_{\alpha \beta , \alpha' \beta'}(12)\psi_{\alpha' \beta'}(12)
\nonumber\\
&&+\sum_{\alpha'}V_{\alpha \beta , \alpha' \alpha'} (12)
\psi_{\alpha' \alpha'}(12) = E \psi_{\alpha \beta}(12)
\label{2.10}
\end{eqnarray}

 Again introducing the arbitrary but fixed ordering into the
 different particle
labels  leads  to

\begin{eqnarray}
&&(\frac{p_{1}^{2}}{2m_{\alpha}} + \frac{p_{2}^{2}}{2m_{\beta}} + m_{\alpha}
 + m_{\beta})
\psi_{\alpha \beta}(12)+\sum_{\alpha' < \beta'}
V_{\alpha \beta , \alpha' \beta'}^{tot}(12)\psi_{\alpha' \beta'}(12)
\nonumber\\
&&+\sum_{\alpha'}V_{\alpha \beta , \alpha' \alpha'} (12)
\psi_{\alpha' \alpha'}(12) = E \psi_{\alpha \beta}(12)
\label{2.11}
\end{eqnarray}

Using Eq.~(2.11) we introduced for $\alpha '\neq\beta ß$
the "total"-potential
 leading from $\alpha\beta$ to $\alpha '\beta '$ 
which is the sum of a
 "direct" potential and an "exchange" potential

\begin{equation}
V_{\alpha \beta , \alpha' \beta'}^{tot}(12) \equiv V_{\alpha
 \beta , \alpha' \beta'}(12)-V_{\alpha \beta , \beta' \alpha'}(12)P_{12}
\label{2.12}
\end{equation}

The application  of Eq. (2.15) to $|\, \alpha \, \beta>=\{
|\, N\Lambda>, \, |\, N\Sigma> \}$
is obvious and will not be written down. For
$|\, \alpha \, \beta>\equiv\{
|\, \Xi N>, \, |\, \Lambda\Lambda>, \, |\, \Sigma\Sigma >, \, 
|\, \Lambda\Sigma >
 \}$
we obtain the set
of four coupled equations

\begin{eqnarray}
(\frac{p_{1}^{2}}{2m_{\Xi}}&& + \frac{p_{2}^{2}}{2m_{N}} + m_{\Xi}
 + m_{N} -E)\psi_{\Xi N}(12) + V_{\Xi N,\Xi N}(12)\psi_{\Xi N}(12)
\nonumber\\
&&+V_{\Xi N,\Lambda \Sigma}^{tot}(12)\psi_{\Lambda \Sigma}(12)
+V_{\Xi N,\Lambda \Lambda}(12)\psi_{\Lambda \Lambda}(12)
+V_{\Xi N,\Sigma \Sigma}(12)\psi_{\Sigma \Sigma}(12)=0
\\
\nonumber\\
(\frac{p_{1}^{2}}{2m_{\Lambda}}&& + \frac{p_{2}^{2}}{2m_{\Sigma}} 
+ m_{\Lambda} + m_{\Sigma} -E)\psi_{\Lambda \Sigma}(12) 
+ V_{\Lambda \Sigma,\Lambda \Sigma}^{tot}(12)\psi_{\Lambda \Sigma}(12)
\nonumber\\
&&+V_{\Lambda \Sigma,\Xi N}^{tot}(12)\psi_{\Xi N}(12)
+V_{\Lambda \Sigma,\Sigma \Sigma}(12)\psi_{\Sigma \Sigma}(12)=0
\\
(\frac{p_{1}^{2}}{2m_{\Lambda}}&& + \frac{p_{2}^{2}}{2m_{\Lambda}}
+2m_{\Lambda}-E)\psi_{\Lambda \Lambda}(12)
 + V_{\Lambda \Lambda,\Xi N}^{tot}(12)\psi_{\Xi N}(12)
\nonumber\\
&&+V_{\Lambda \Lambda,\Sigma \Sigma}(12)
\psi_{\Sigma \Sigma}(12)+V_{\Lambda \Lambda,\Lambda \Lambda}(12)
\psi_{\Lambda \Lambda}(12)=0
\\
(\frac{p_{1}^{2}}{2m_{\Sigma}}&& + \frac{p_{2}^{2}}{2m_{\Sigma}}
+2m_{\Sigma}-E)\psi_{\Sigma \Sigma}(12)
+V_{\Sigma \Sigma,\Xi N}^{tot}(12)
\psi_{\Xi N}(12)+V_{\Sigma \Sigma,\Lambda \Sigma}^{tot}(12)
\psi_{\Lambda \Sigma}(12)
\nonumber\\
&&+V_{\Sigma \Sigma,\Lambda \Lambda}(12)\psi_{\Lambda \Lambda}(12)
+V_{\Sigma \Sigma,\Sigma \Sigma}(12)\psi_{\Sigma \Sigma}(12)=0
\end{eqnarray}

 This set has to be solved under the condition that
 $\psi_{\Lambda\Lambda} (12)$ and
 $\psi_{\Sigma\Sigma} (12)$ are antisymmetric, which
imposes conditions on the wavefunction components with two
different particle species. In [1] we found three coupled equations
regarding however only the coupled $\Xi N - \Lambda\Lambda$ system.
Dropping in Eqs.~(2.17-20) the $\Sigma$ state we  have now two equations.
The second of the three equations
(2.15) from \cite{ours}  has there been introduced 
in order to have a symmetrical
 potential matrix. But it results simply from the first equation
 in that set (2.15) by permuting the two particles
and using the property  (2.13) in \cite{ours}  which is also
given here as  Eq. (2.12). The 
remaining two equations in [1]  are identical to Eq.~(2.17) and
 Eq.~(2.19). The
 interaction
$V_{\Xi N}$ in \cite{ours} includes the exchange term according to
its construction  and is  the same as 
$V^{tot}_{\Xi N ,\, \Xi N}$.
 As is seen from Eq. (2.13) the  total $\Psi$
 contains  both, symmetrical and antisymmetrical
 pieces,
 in the parts of the wavefunction referring to different particles. 

We would like to add that in \cite{ours}  unfortunately some errors
sneaked in.
After Eq.~(2.3) it is said that the variables x or y include isospin
quantum numbers, for instance. This is incorrect, x or y stand only
 for ordinary  space and spin variables. Also the description of 
the examples following Eq.~(2.22) is incorrectly given.

\section{Three-Baryon Systems}

We start off by introducing three-particle states for total strangeness
$S=-1$ and $S=-2$ systems. For instance in case of $S=-1$ they are 
$|\, \Lambda NN>$, $|\, N\Lambda N>$, $|\, NN\Lambda >$,
 $|\, \Sigma NN>$, 
$|\, N\Sigma N>$ and $|\, NN\Sigma>$,
whereas for $S=-2$ one has 15 states composed of
$N$, $\Lambda$, $\Sigma$ and $\Xi$ species. A general state
is denoted by  $|\,\alpha \,\beta\,\gamma>$ where  $\alpha$, $\beta$,
 $\gamma$ run over
the different species.
Of course the sets of states contain only physically allowed
 combinations which conserve charge, strangeness and isospin.
 The completeness relation in particle space reads

\begin{equation}
\sum_{\alpha \,\beta \,\gamma}
 |\,\alpha\,\beta\,\gamma> <\alpha\,\beta\,\gamma |=1
\label{3.1}
\end{equation}

and the fully symmetrised Hamiltonian is

\begin{equation}
H = \sum_{\alpha \beta \gamma}\sum_{\alpha' \beta' \gamma'}
|\alpha \beta \gamma ><\alpha \beta \gamma|H|\alpha' \beta' \gamma'>
<\alpha' \beta' \gamma'|
\label{3.2}
\end{equation}

The eigenstates $\Psi$ of $H$ are then totally antisymmetrical. In what
follows we derive coupled Faddeev equations. The total state $\Psi$ 
will first be decomposed into three Faddeev amplitudes, which due to the
antisymmetry of $\Psi$ can be written as\cite{glockle}

\begin{equation}
\Psi = (1+P)\psi
\label{3.3}
\end{equation}

\noindent
Here $\psi$ is one of the three Faddeev amplitudes and
 $P=P_{12}P_{23}+P_{13}P_{23}$ is the sum
of a cyclical and an anticyclical permutation of three objects.
If $\psi$ is antisymmetrical  under exchange of two particles
then $\Psi$ given in Eq.~(3.3) is fully antisymmetrical.
That one Faddeev amplitude is based on one pair interaction
contained in $H$, say
between particles 2 and 3, and is defined as 

\begin{equation}
\psi = G_{0}V(23)\Psi
\label{3.4}
\end{equation}

\noindent 
where $G_0$ is the free three-baryon propagator.
Consequently it obeys the equation

\begin{equation}
\psi =  G_{0}V(23)\psi + G_{0}V(23)P\psi
\label{3.5}
\end{equation}

\noindent
Since the symmetry condition for $\psi$ is a simple one,
namely to be antisymmetrical under exchange of particles 2 and 3,
we can simply
introduce a complete set of particle states which are antisymmetrical
and symmetrical under exchange of particles 2 and 3:

\begin{equation}
|\chi_{\alpha \beta \gamma}^{a}> \equiv \frac{1}{\sqrt{2}}
(|\alpha \beta \gamma>-|\alpha \gamma \beta>)  \,\,\,\,\,\,\,\, 
\beta \neq \gamma
\label{3.6}
\end{equation}
\begin{equation}
|\chi_{\alpha \beta \gamma}^{s}> \equiv \frac{1}{\sqrt{2}}
(|\alpha \beta \gamma>+|\alpha \gamma \beta>)  \,\,\,\,\,\,\,\, 
\beta \neq \gamma
\label{3.7}
\end{equation}
\begin{equation}
|\chi_{\alpha \beta \beta}^{s}> \equiv |\alpha \beta \beta>
\label{3.8}
\end{equation}
 The general form

\begin{equation}
|\psi>=
\sum_{\alpha \,\beta\,\gamma}
 |\, \alpha \, \beta\,\gamma> <\alpha \, \beta \,\gamma\, |\psi>
\equiv\sum_{\alpha \,\beta,\gamma} |\, \alpha \, \beta\,\gamma>
\psi_{\alpha \,\beta\,\gamma} 
\end{equation}

\noindent
can then be rewritten as

\begin{equation}
|\psi>=\sum_{\alpha \,\beta}
 |\chi^s_{\alpha\beta\beta}> \psi_{\alpha\beta\beta}
+
\sum_{\alpha}\sum_{\beta\neq\gamma}\frac{1}{\sqrt{2}}
(|\chi_{\alpha \beta\gamma}^{a}>
+|\chi_{\alpha\beta\gamma}^{s} >)
 \psi_{\alpha\beta\gamma}
\end{equation}

\noindent
Imposing antisymmetry under exchange of particles 2 and 3 leads
 to

\begin{equation}
P_{23}\psi_{\alpha \beta\gamma}=-\psi_{\alpha\gamma\beta}
\end{equation}

\noindent
and after some inspection to

\begin{eqnarray}
\psi&=& \sum_{ \alpha} \sum_{ \beta < \gamma}
 |\chi_{ \alpha \beta \gamma}^{a}> \frac{1}{ \sqrt{2}}(1+P_{23}) 
\psi_{ \alpha \beta \gamma}+ \sum_{ \alpha} \sum_{ 
\beta < \gamma} |\chi_{ \alpha \beta \gamma}^{s} >
\frac{1}{ \sqrt{2}}(1-P_{23}) \psi_{ \alpha \beta \gamma}
\nonumber\\
&+& \sum_{ \alpha \beta}| \chi_{ \alpha \beta \beta}^{s}> 
\psi_{ \alpha \beta \beta}
\label{3.12}
\end{eqnarray}

The set of coupled equations for $\psi_{\alpha\beta\gamma}$
is based on Eq.~(3.5), which when projected reads

\begin{equation}
\psi_{ \alpha \beta \gamma}=G_{0}^{ \alpha \beta \gamma}
(<\alpha \beta \gamma |V(23)| \psi>+< \alpha \beta \gamma |V(23)P| \psi>)
\label{3.13}
\end{equation}

\noindent
Clearly the free propagators $G_0^{\alpha\beta\gamma}$ are
diagonal and depend on the particle
species entering through their masses. 
The order $\alpha\beta\gamma$ corresponds of course
to particles numbered 1,2 and 3.
The first matrix element on the right  hand side
of Eq.~(3.13) is easily evaluated as

\begin{equation}
\begin{array}{l}
< \alpha \beta \gamma |V| \psi>
\\
= \displaystyle{\sum_{ \beta ' \gamma '}} 
\bar{\delta}_{ \beta ' \gamma '}V_{ \beta \gamma , \beta ' \gamma '}(23)
< \alpha \beta ' \gamma '| \psi>
+ \displaystyle{\sum_{ \beta '}}
V_{ \beta \gamma , \beta ' \beta '}(23)< \alpha \beta ' \beta '| \psi>
\\
= \displaystyle{\sum_{ \beta '< \gamma '}}
(V_{ \beta \gamma , \beta ' \gamma '}(23)
-V_{ \beta \gamma , \gamma ' \beta '}(23)P_{23}) 
\psi_{ \alpha \beta ' \gamma '}
+ \displaystyle{\sum_{ \beta '}}
V_{ \beta \gamma , \beta ' \beta '}(23) \psi_{ \alpha \beta ' \beta '}
\\
\equiv \displaystyle{\sum_{ \beta '< \gamma '}}
V_{ \beta \gamma , \beta ' \gamma '}^{tot}(23) 
\psi_{ \alpha \beta ' \gamma '}+ \displaystyle{\sum_{ \beta '}}
V_{ \beta \gamma , \beta ' \beta '}(23) \psi_{ \alpha \beta ' \beta '}
\end{array}
\label{3.14}
\end{equation}

\noindent
Note we used the property (3.11) and introduced the convenient notation
$\bar{\delta}_{\beta\gamma}\equiv 1-\delta_{\beta\gamma}$.

In order to evaluate the second matrix element on the right hand
side of Eq.~(3.13) we need $P\psi$.  Using (3.9) one finds after a
simple algebra

\begin{equation}
<\alpha \beta \gamma|P|\psi >=P_{12}P_{23}
\psi_{\beta \gamma \alpha} +P_{13}P_{23}
\psi_{\gamma \alpha \beta} 
\label{3.17}
\end{equation}

\noindent
This then leads easily to

\begin{eqnarray}
<\alpha \beta \gamma|VP|\psi >
&=&
{\sum_{\beta'<\gamma'}}
V_{\beta \gamma,\beta' \gamma'}^{tot}(23)\{P_{12}P_{23}
\psi_{\beta' \gamma' \alpha}
+P_{13}P_{23}\psi_{\gamma' \alpha \beta'}
\}
\nonumber
\\
&&+
{\sum_{\beta'}}
V_{\beta \gamma, \beta' \beta'}(23)(1-P_{23})P_{13}P_{23}
\psi_{\beta' \alpha \beta '}
\label{3.18}
\end{eqnarray}

\noindent
Here we used the property 

\begin{equation}
P_{23}P\psi =-P\psi
\end{equation}

\noindent
which is evident from Eq. (3.12).
Put together we end up with the following set of coupled equations 
for the amplitudes $\psi_{\alpha\beta\gamma}$

\begin{eqnarray}
\psi_{\alpha \beta \gamma} 
= G_{0}^{\alpha \beta \gamma} 
(\sum_{\beta' < \gamma'} V_{\beta \gamma,\beta' \gamma'}^{tot}(23)
\psi_{\alpha \beta' \gamma'}
+ 
\sum_{\beta'} V_{\beta \gamma,\beta' \beta'}(23)
\psi_{\alpha \beta' \beta'} )
\nonumber\\
+G_{0}^{\alpha \beta \gamma} \{
\sum_{\beta' < \gamma'} V_{\beta \gamma,\beta' \gamma '}^{tot}(23)
 (P_{12}P_{23}\psi_{\beta' \gamma' \alpha}
 +P_{13}P_{23}\psi_{\gamma' \alpha \beta'}
 )
\nonumber\\
+
\sum_{\beta'}V_{\beta \gamma,\beta' \beta'}(23)(1-P_{23})
P_{13}P_{23}\psi_{\beta ' \alpha \beta '} \}
\label{3.19}
\end{eqnarray}

We cast it into the following matrix form

\begin{equation}
\tilde{\psi}=\tilde{G_{0}}\tilde{V}\tilde{\psi}
+\tilde{G_{0}}\tilde{V}\tilde{R}
\label{3.20}
\end{equation}

\noindent
where $\tilde G_0$ is diagonal and contains the various
free propagators $G_0^{\alpha\beta\gamma}$.
The matrix $\tilde V$ is composed of all non vanishing 
diagonal and transition potentials and $\tilde R$ of the
 $\psi_{\alpha\beta\gamma}$
amplitudes with permutation operators in front.
Thus the two first terms on the right hand side of Eq. (3.18)
define $\tilde V$ and all the rest $\tilde R$.
In a standard manner one can sum up $\tilde V$ to infinite order leading
 to

\begin{equation}
\tilde{\psi}=\tilde{\psi^{0}}+\tilde{G_{0}}\tilde{t}\tilde{R}
\label{3.21}
\end{equation}

where $\tilde t$ obeys the matrix Lippmann-Schwinger equation 

\begin{equation}
\tilde{t}=\tilde{V}+\tilde{V}\tilde{G_{0}}\tilde{t}
\label{3.22}
\end{equation}

If one regards a scattering process initiated by some particle
hitting a two-baryon bound state then the driving term $\tilde{\psi^0}$
 contains
a nonzero entry in the corresponding row. For a bound system
 $\tilde{\psi^0}$ is absent.

As a first example let us regard the strangeness $-1$ system
consisting of the $N$, $\Lambda$ and $\Sigma$ particles.
Let us group the channels in the following manner:
$\Lambda NN$, $\Sigma NN$, $N\Lambda N$, $N\Sigma N$.
Then the matrix $\tilde V$ has the form 

\begin{equation}
\tilde{V}=
\left(
\begin{array}{cccc}
V_{NN} & 0 & 0 & 0\\
0 & V_{NN} & 0 & 0\\
0 & 0 & V_{\Lambda N , \Lambda N}^{tot} & V_{\Lambda N , 
\Sigma N}^{tot}\\
0 & 0 & V_{\Sigma N , \Lambda N}^{tot} & V_{\Sigma N , \Sigma N}^{tot}
\end{array}
\right)
\label{3.23}
\end{equation}

\noindent
Consequently the Lippman-Schwinger equation (3.21) will decay into 
two uncoupled equations ($Y=\Lambda$ and $\Sigma$)

\begin{equation}
t_{NN}^{Y}=V_{NN}+N_{NN}G_{0}^{YNN}t_{NN}^{Y}
\label{3.24}
\end{equation}

\noindent
and one set of two coupled equations 

\begin{equation}
\tilde{t}=
\left(
\begin{array}{cc}
V_{ \Lambda N , \Lambda N}^{tot} & V_{ \Lambda N , \Sigma N}^{tot} \\
V_{ \Sigma N , \Lambda N}^{tot} & V_{ \Sigma N , \Sigma N}^{tot}
\end{array}
\right)
+
\left(
\begin{array}{cc}
V_{ \Lambda N , \Lambda N}^{tot} & V_{ \Lambda N , \Sigma N}^{tot} \\
V_{ \Sigma N , \Lambda N}^{tot} & V_{ \Sigma N , \Sigma N}^{tot}
\end{array}
\right)
\left(
\begin{array}{cc}
G_{0}^{N \Lambda N} & 0 \\
0 & G_{0}^{N \Sigma N}
\end{array}
\right)
\tilde{t}
\label{3.25}
\end{equation}
 
We thus end up with the following four coupled Faddeev equations

\begin{equation}
\psi_{ \Lambda NN}= \psi_{ \Lambda d}^{0}+G_{0}^{ \Lambda NN}
t_{NN}^{ \Lambda}(1-P_{23})P_{13}P_{23} \psi_{N \Lambda N}
\label{3.26}
\end{equation}
\begin{equation}
\psi_{ \Sigma NN}= \psi_{ \Sigma d}^{0}+G_{0}^{ 
\Sigma NN}t_{NN}^{ \Sigma}(1-P_{23})P_{13}P_{23} \psi_{N \Sigma N}
\label{3.27}
\end{equation}
\begin{equation}
\left(
\begin{array}{c}
\psi_{N \Lambda N} \\
\psi_{N \Sigma N}
\end{array}
\right)
=
\left(
\begin{array}{cc}
G_{0}^{N \Lambda N} & 0 \\
0 & G_{0}^{N \Sigma N}
\end{array}
\right)
\tilde{t}
\left(
\begin{array}{c}
-P_{13} \psi_{N \Lambda N}+P_{12}P_{23} \psi_{ \Lambda NN} \\
-P_{13} \psi_{N \Sigma N}+P_{12}P_{23} \psi_{ \Sigma NN}
\end{array}
\right)
\label{3.28}
\end{equation}

\noindent
The driving terms in Eqs.(3.25) and (3.26) occur if we regard
$\Lambda d$ or $\Sigma d$ scattering. Of course only the one 
or the other will be nonzero.

It is an easy exercise
using appropriate permutations to verify that the homogeneous set
corresponding to  Eqs.~(3.25)-(3.27)  is 
identical to the set
solved numerically in \cite{hypt}.
The total hypertriton state $\Psi$ is then given by (3.3). This can
 be projected 
into the various particle species.

The case of strangeness $S=-2$ is more complex.
We use the following arbitrary but fixed grouping of 
particle states:
$\Xi NN$, $N\Xi N$, $N\Lambda \Sigma$, $N\Lambda\Lambda$, 
$N\Sigma\Sigma$, $\Lambda N\Lambda$, $\Lambda N\Sigma$,
$\Sigma N\Lambda$ and $\Sigma N\Sigma$.
Note the remaining six groupings can be related to the nine
given ones with the help of Eq. (3.11).
Then we can read off from Eq.(3.18) a $9\times 9$ potential matrix.
For the sake of an easier notation we drop the $\Sigma$ particle
and find

\begin{equation}
\tilde{V}=
\left(
\begin{array}{cccc}
V_{NN} & 0 & 0 & 0 \\
0 & V_{ \Xi N , \Xi N} & V_{ \Xi N , \Lambda \Lambda} & 0 \\
0 & V_{ \Lambda \Lambda , \Xi N}^{tot} & V_{ \Lambda \Lambda} & 0 \\
0 & 0 & 0 & V_{N \Lambda , N \Lambda}^{tot}
\end{array}
\right)
\label{3.29}
\end{equation}

Again the matrix Lippman-Schwinger equation for $\tilde t$ decays
in a corresponding manner and we find the following four 
coupled Faddeev equations.

\begin{equation}
\psi_{ \Xi NN}= \phi_{ \Xi d}+G_{0}^{ \Xi NN}
t_{NN}(1-P_{23})P_{13}P_{23} \psi_{N \Xi N}
\label{3.30}
\end{equation}
\begin{equation}
\left(
\begin{array}{c}
\psi_{N \Xi N} \\
\psi_{N \Lambda \Lambda}
\end{array}
\right)
=
\left(
\begin{array}{cc}
G_{0}^{N \Xi N} & 0 \\
0 & G_{0}^{N \Lambda \Lambda}
\end{array}
\right)
\tilde{t}
\left(
\begin{array}{c}
P_{12}P_{23} \psi_{ \Xi NN}-P_{13} \psi_{N \Xi N} \\
(1-P_{23})P_{13}P_{23} \psi_{ \Lambda N \Lambda}
\end{array}
\right)
\label{3.31}
\end{equation}
\begin{equation}
\psi_{ \Lambda N \Lambda}=G_{0}^{ \Lambda N \Lambda}t_{N\Lambda}
(P_{12}P_{23}\psi_{N \Lambda\Lambda}
  -P_{13}\psi_{\Lambda N \Lambda} )
\label{3.32}
\end{equation}

\noindent
These four equations are identical to the ones presented
in \cite{ours}.
In order to demonstrate that identity it is again necessary to apply
appropriate permutations. This is left to the reader.
For later use it is helpful to write that set (3.29)-(3.31) explicitly
into the form (3.20). One easily finds 

\begin{equation}
\begin{array}{c}
\left(
\begin{array}{c}
\psi_{ \Xi NN} \\
\psi_{N \Xi N} \\
\psi_{N \Lambda \Lambda} \\
\psi_{ \Lambda N \Lambda}
\end{array}
\right)
=
\left(
\begin{array}{c}
\phi_{\Xi d} \\
0 \\
0 \\
0
\end{array}
\right)
\hfill
\\
\\
+
\left(
\begin{array}{cccc}
G_{0}^{ \Xi NN} & 0 & 0 & 0 \\
0 & G_{0}^{N \Xi N} & 0 & 0 \\
0 & 0 & G_{0}^{N \Lambda \Lambda} & 0 \\
0 & 0 & 0 & G_{0}^{ \Lambda N \Lambda}
\end{array}
\right)
\left(
\begin{array}{cccc}
 t_{NN} & 0 & 0 & 0 \\
  0 &t_{ \Xi N , \Xi N} &  t_{ \Xi N , \Lambda \Lambda}  & 0 \\
  0 &t_{ \Lambda \Lambda , \Xi N} &  t_{ \Lambda \Lambda , \Lambda \Lambda} & 0 \\
0 & 0 & 0 & t_{N \Lambda , N \Lambda}
\end{array}
\right)
\hfill
\\
\\
\left(
\begin{array}{cccc}
0 & (1-P_{23})P_{13}P_{23} & 0 & 0 \\
P_{12}P_{23} & -P_{13} & 0 & 0 \\
0 & 0 & 0 & (1-P_{23})P_{13}P_{23} \\
0 & 0 & P_{12}P_{23} & -P_{13}
\end{array}
\right)
\left(
\begin{array}{c}
\psi_{ \Xi NN} \\
\psi_{N \Xi N} \\
\psi_{N \Lambda \Lambda} \\
\psi_{ \Lambda N \Lambda}
\end{array}
\right)\hfill
\end{array}
\end{equation}

\noindent 
This identifies explicitly the term $\tilde t\tilde R$ as
$\tilde t\tilde P\tilde \psi$, where the matrix of permutation
 operators $\tilde P$ can be read off from Eq. (3.32). Then Eq.(3.20)
 can be expressed as

\begin{equation}
\tilde{\psi}=\tilde{\psi^{0}}+\tilde{G_{0}}\tilde{t}\tilde{P}\tilde \psi
\end{equation}

In case of  a bound three-body system the driving term
$\phi_{\Xi d}$, which is a product of a momentum eigenstate
for the $\Xi$ particle and the deuteron state, is of course 
absent.
It is not yet known for sure whether that interesting system fluctuating
between $\Xi NN$, $\Lambda\Lambda N$, $\Lambda\Sigma N$ and 
$\Sigma\Sigma N$                          
and which is described by a set of nine coupled Faddeev equations 
is bound or exists at least as a resonance.

\section{Breakup processes in $\Xi\, d$ Scattering}

In \cite{afnan} it was proposed that the process 
$\Xi d\rightarrow\Lambda\Lambda N$
might be a source of information for the 
$\Lambda\Lambda$ scattering length.
If the $\Lambda\Lambda$ system has a virtual state
close to zero relative energy one has to expect an enhancement
in the breakup configuration, when the two $\Lambda$'s leave with equal
laboratory momenta. In \cite{afnan} it was found in a numerical study
using model forces that a destructive interference
between different production amplitudes diminishes that enhancement.
The strength of that suppression turned out to be linked to the
$\Lambda\Lambda -\Xi N$
coupling and is therefore of great theoretical interest. As already
 touched
upon in the introduction the formalism in \cite{afnan} is based on the
 AGS equations
for distinguishable particles and antisymmetrisation is  attached
through the  antisymmetrised asymptotic initial and 
final channel states. Obviously it is
desirable to have a straightforward derivation of the appropriate
scattering formalism. This is achieved in the form of the coupled set
of Faddeev equations (3.29)-(3.31)
 for the four independent Faddeev amplitudes. 
(In case of an additional $\Sigma$-admixture one has nine independent 
amplitudes.)
They describe the elastic process

\begin{equation}
\Xi+d \rightarrow \Xi+d
\label{4.1}
\end{equation}

\noindent
and the various breakup processes

\begin{equation}
\Xi+d \to
\left\{
\begin{array}{c}
\Xi+N+N \\
\Lambda+ \Lambda+N \\
N+ \Lambda+ \Sigma \\
N+ \Sigma+ \Sigma
\end{array}
\right.
\label{4.2}
\end{equation}

It is a standard procedure \cite{glockle,polonica,physrep,zeit} 
to extract the various
 partial breakup
amplitudes from each individual kernel part of the set (3.29)-(3.31)
 or in general matrix
notation from the kernel of (3.20).
One regards the asymptotic behavior in configuration space and finds
 the partial
breakup amplitudes to be 

\begin{equation}
\tilde{T} \equiv \tilde{t} \tilde{R}
\label{4.3}
\end{equation}

\noindent
Since $\tilde R=\tilde P\tilde\psi $ 
and  

\begin{equation}
\tilde{ \psi}= \tilde{ \psi^{0}}+ \tilde{G_{0}} \tilde{T}
\label{4.4}
\end{equation}

\noindent
one arrives at 
\begin{equation}
\tilde{T}=\tilde t\tilde P\tilde{\psi^0}+\tilde t\tilde P\tilde{G_{0}}\tilde T
\end{equation}

\noindent
which is  a closed set of equations for all partial
breakup amplitudes.
Again for the sake of a simple notation we regard the set
 (3.29)-(3.31) combined in Eq.(3.32) without $\Sigma$
and easily find that Eq. (4.5) takes the explicit form:

\begin{equation}
\begin{array}{c}
\left(
\begin{array}{c}
T_{ \Xi NN} \\
T_{N \Xi N} \\
T_{N \Lambda \Lambda} \\
T_{ \Lambda N \Lambda}
\end{array}
\right)
=
\left(
\begin{array}{c}
0 \\
t_{ \Xi N , \Xi N}P_{12}P_{23}\phi_{\Xi d} \\
t_{ \Lambda \Lambda , \Xi N}P_{12}P_{23}\phi_{\Xi d} \\
0
\end{array}
\right)
\hfill
\\
\\
+\left(
\begin{array}{cccc}
0 & t_{NN}(1-P_{23})P_{13}P_{23} & 0 & 0 \\
t_{ \Xi N , \Xi N}P_{12}P_{23} & -t_{ \Xi N , \Xi N}P_{13} & 0 
& t_{ \Xi N , \Lambda \Lambda}(1-P_{23})P_{13}P_{23} \\
t_{ \Lambda \Lambda , \Xi N}P_{12}P_{23} & 
-t_{ \Lambda \Lambda , \Xi N}P_{13} & 0 
& t_{ \Lambda \Lambda , \Lambda \Lambda}(1-P_{23})P_{13}P_{23} \\
0 & 0 & t_{N \Lambda , N \Lambda}P_{12}P_{23} 
& -t_{N \Lambda , N \Lambda}P_{13}
\end{array}
\right)
\\
\\
\left(
\begin{array}{cccc}
G_{0}^{ \Xi NN} & 0 & 0 & 0 \\
0 & G_{0}^{N \Xi N} & 0 & 0 \\
0 & 0 & G_{0}^{N \Lambda \Lambda} & 0 \\
0 & 0 & 0 & G_{0}^{ \Lambda N \Lambda}
\end{array}
\right)
\left(
\begin{array}{c}
T_{ \Xi NN} \\
T_{N \Xi N} \\
T_{N \Lambda \Lambda} \\
T_{ \Lambda N \Lambda}
\end{array}
\right)\hfill
\end{array}
\end{equation}

The full breakup amplitudes result from projecting the full
wavefunction given in Eq. (3.3) onto the
 various
breakup configurations. Using (3.15) one finds for instance

\begin{equation}
< \Xi NN| \Psi>=< \Xi NN|(1+P)| \psi>= \psi_{ \Xi NN}
+(1-P_{23})P_{13}P_{23} \psi_{N \Xi N}
\label{4.6}
\end{equation}

\noindent
Consequently the full breakup amplitude into the channel $\Xi NN$
is

\begin{equation}
U_{ \Xi NN}^{0}=T_{ \Xi NN}+(1-P_{23})P_{13}P_{23}T_{N \Xi N}
\label{4.7}
\end{equation}

\noindent
Similarly we get for the breakup process into $\Lambda N\Lambda$

\begin{equation}
U_{ \Lambda N\Lambda }^{0}=(1-P_{13})T_{ \Lambda N \Lambda}
+P_{12}P_{23}T_{N \Lambda \Lambda}
\label{4.8}
\end{equation}

Note both amplitudes have the appropriate antisymmetry for $NN$ 
and $\Lambda\Lambda$, respectively as guaranteed by the properties
$P_{23}T_{\Xi NN}=-T_{\Xi NN}$ and 
$P_{23}T_{N\Lambda\Lambda}=-T_{N\Lambda\Lambda}$.
All further projections lead to no new breakup amplitudes.
If we would have projected for instance in case
that in addition the $\Sigma$ is included  for the $\Lambda\Sigma N$
breakup 
the resulting amplitude $<\Lambda\Sigma N\,|\, \Psi>$ 
is connected to $<N\Lambda\Sigma \,|\, \Psi>$
by $P_{23}P_{13} <\Lambda\Sigma N\,|\, \Psi>=
<N\Lambda\Sigma \,|\, \Psi>$ which tells that only the 
particles have been numbered differently.

Since after all those $YN$ and $YY$ forces are weaker than
the $NN$ force it is not unreasonable to look into
a multiple scattering series expansion
 of the set (4.5).
It might even converge if in the future it will turn out that there is
no strangeness $-2$ three-baryon  bound state  nor even a low
 lying three-body resonance.
The multiple scattering series results by iterating the set(4.6).
In lowest and next to lowest order  in the $t$-operators 
one gets

\begin{equation}
T_{N \Xi N}=-t_{ \Xi N , \Xi N}(23)P_{12} \phi_{ \Xi d}
+t_{ \Xi N , \Xi N}(23)P_{13}G_{0}^{N \Xi N}t_{ \Xi N , \Xi N}(23)P_{12} 
\phi_{ \Xi d}
\label{4.9}
\end{equation}
\begin{equation}
T_{N \Lambda \Lambda}=-t_{ \Lambda \Lambda , \Xi N}(23)P_{12} 
\phi_{ \Xi d}+t_{ \Lambda \Lambda , \Xi N}(23)P_{13}G_{0}^{N \Xi N}
t_{ \Xi N , \Xi N}(23)P_{12} \phi_{ \Xi d}
\label{4.10}
\end{equation}
\begin{equation}
T_{ \Lambda N \Lambda}=-t_{N \Lambda , N \Lambda}(23)P_{12}P_{23}
G_{0}^{N \Lambda \Lambda}t_{ \Lambda \Lambda , \Xi N}(23)P_{12} 
\phi_{ \Xi d}
\label{4.11}
\end{equation}
\begin{equation}
T_{ \Xi NN}=t_{NN}(23)(1-P_{23})P_{12}G_{0}^{N \Xi N}
t_{ \Xi N , \Xi N}(23)P_{12} \phi_{ \Xi d}
\label{4.12}
\end{equation}

\noindent
Using Eqs. (4.8) and (4.9) we arrive at the two breakup
 amplitudes to second order in the $t$-matrices:

\begin{eqnarray}
U_{ \Xi NN}^{0}&=& t_{ \Xi N,\, \Xi N}(12)\,\phi_{\Xi d}
                  +t_{ \Xi N,\, \Xi N}(13)\,\phi_{\Xi d}
\hfill\nonumber\\
  &+& t_{ \Xi N,\, \Xi N}(12)\, G_0^{ \Xi NN}\,
                 t_{ \Xi N,\, \Xi N}(13)\,\phi_{\Xi d}
    + t_{ \Xi N,\, \Xi N}(13)\, G_0^{ \Xi NN}\,
                 t_{ \Xi N,\, \Xi N}(12)\,\phi_{\Xi d}
\hfill\nonumber\\
  &+& t_{N N}(23)\, G_0^{ \Xi NN}\,
                 t_{ \Xi N,\, \Xi N}(13)\,\phi_{\Xi d}
    + t_{N N}(23)\, G_0^{ \Xi NN}\,
                 t_{ \Xi N,\, \Xi N}(12)\,\phi_{\Xi d}
\hfill \\ 
U_{ \Lambda N\Lambda}^{0}&=&t_{ \Lambda\Lambda ,\, \Xi N}(13)\,\phi_{\Xi d}
\hfill\nonumber\\
  &+& t_{N\Lambda , \, N\Lambda}(23)\, G_0^{ \Lambda N\Lambda}\,
          t_{ \Lambda\Lambda ,\, \Xi N}(13)\,\phi_{\Xi d}
    + t_{N\Lambda , \, N\Lambda}(21)\, G_0^{ \Lambda N\Lambda}\,
          t_{ \Lambda\Lambda ,\, \Xi N}(13)\,\phi_{\Xi d}
\hfill\nonumber\\
   &+& t_{ \Lambda\Lambda ,\, \Xi N}(13)\, G_0^{ \Xi NN}\,
                 t_{ \Xi N,\, \Xi N}(12)\,\phi_{\Xi d}
\hfill
\end{eqnarray}

\noindent
In arriving at Eq. (4.15) we used

\begin{equation}
  P_{23} t_{ \Lambda\Lambda ,\, \Xi N}(23)
              =-t_{ \Lambda\Lambda ,\, \Xi N}(23)
\end{equation}

\noindent
which is direct consequence of the definition (2.16) 
of $V_{\alpha\beta ,\alpha '\beta '}^{tot}$
and the corresponding Lippmann-Schwinger equation  
(see Appendix).
The antisymmetry of $U^0_{\Xi NN}$ in the two nucleons,
 $P_{23} U^0_{\Xi NN}=- U^0_{\Xi NN}$, can easily be seen 
to be guaranteed by the antisymmetry of the deuteron state.
In case of $ U^0_{\Lambda N\Lambda}$ the antisymmetry
 $P_{13} U^0_{\Lambda N\Lambda}=- U^0_{\Lambda N\Lambda}$
is due to the property (4.16).
We exhibit the content of Eqs. (4.14) and (4.15)
 graphically in Figs. 1 and 2.
With the techniques available nowadays (see for instance
\cite{physrep} ) it is feasible to evaluate these processes
reliably using modern forces. 

\section{Elastic $\Xi d$ Scattering-Generalized AGS Equations
}

The two sets of four coupled Faddeev equations (3.25-3.27)
and (3.29)-(3.31) or more generally (3.33) also define the
asymptotic behavior in bound state channels. As known up
to now there is only one baryon-baryon bound state, the deuteron.
Using the well known relation

\begin{equation}
(E-H_{0})^{-1}t_{NN}=(E-H_{0}- V_{NN})^{-1} V_{NN}
\label{5.1}
\end{equation}

\noindent
one can extract easily the asymptotic behavior in configuration
space in the deuteron channel\cite{zeit}. For instance the elastic
 amplitude
in $\Xi d$ scattering extracted from Eq. (3.29) is

\begin{equation}
U_ {\Xi d}=< \phi '_ {\Xi d}| V_{NN}(23)(1-P_{23})P_{13}P_{23}| 
\psi_{N \Xi N}>
\label{5.2}
\end{equation}

\noindent
where the prime on the channel state $\phi '_{\Xi d}$  indicates that
 the momentum
of the outgoing $\Xi$-particle is independent of the momentum of
the incoming $\Xi$-particle.Inserting the right hand side of the Faddeev
 equation
from Eq. (3.30) for  $\psi_{N\Xi N}$ and using
the definition of $T_{N\Xi N}$ 
 as expressed in Eq. (4.4)
one finds

\begin{equation}
U_{\Xi d}=< \phi '_ {\Xi d}| V_{NN}(23)(1-P_{23})P_{13}P_{23}
G_{0}^{N \Xi N}T_{N \Xi N}>
\label{5.3}
\end{equation}

\noindent
Thus the elastic amplitude can be determined by quadrature from the
partial breakup amplitude $T_{N\Xi N}$.

Though for practical calculations working with general forces the
Faddeev equation for the $T$'s  are most appropriate\cite{physrep} it is
 interesting
from a formal point of view to derive Faddeev equations for transition
operators describing elastic and rearrangement scattering. 
Using forces of finite rank they are also a good starting point.

In the case
of the outgoing $\Xi d$ channel we extracted a state

\begin{equation}
U '_{ \Xi d \Xi d} \equiv V_{NN}(23)(1-P_{23})P_{13}P_{23} 
\psi_{N \Xi N}
\label{5.4}
\end{equation}

\noindent
which under action from the left by $<\phi '_{\Xi d} \, |$
gives the elastic transition
amplitude (see Eq. (5.2).
In the elastic amplitude
 on the energy shell we apparently can  replace $V_{NN}$ by
$(G_0^{\Xi NN})^{-1}=(E-H_0^{\Xi NN})$.                        
Let us therefore define

\begin{equation}
U_{ \Xi d \Xi d} \equiv(G_{0}^{ \Xi NN})^{-1}(1-P_{23})P_{13}P_{23}
 \psi_{N \Xi N}
\label{5.5}
\end{equation}

Analogously we define

\begin{equation}
\left(
\begin{array}{c}
U_{N \Xi N , \Xi d} \\
U_{N \Lambda \Lambda , \Xi d}
\end{array}
\right)
\equiv
\left(
\begin{array}{cc}
(G_{0}^{N \Xi N})^{-1} & 0 \\
0 & (G_{0}^{N \Lambda \Lambda})^{-1} 
\end{array}
\right)
\left(
\begin{array}{c}
P_{12}P_{23} \psi_{ \Xi NN}-P_{13} \psi_{N \Xi N} \\
(1-P_{23})P_{13}P_{23} \psi_{ \Lambda N \Lambda} 
\end{array}
\right)
\label{5.6}
\end{equation}

\begin{equation}
U_{ \Lambda N \Lambda , \Xi d} \equiv (G_{0}^{ \Lambda N \Lambda})^{-1}
(P_{12}P_{23} \psi_{N \Lambda \Lambda}-P_{13} \psi_{ \Lambda N \Lambda})
\label{5.7}
\end{equation}
 
\noindent
These definitions together with the coupled set (3.29)-(3.31)
will lead to a closed set  of equations for the four quantities $U$.
It is advisable to write the four defining equations (5.5)-(5.7)
into a 
matrix form

\begin{equation}
\begin{array}{c}
\tilde{U} \equiv
\left(
\begin{array}{c}
U_{ \Xi d , \Xi d} \\
U_{N \Xi N , \Xi d} \\
U_{N \Lambda \Lambda , \Xi d} \\
U_{ \Lambda N \Lambda , \Xi d}
\end{array}
\right)
=
\left(
\begin{array}{cccc}
(G_{0}^{ \Xi NN})^{-1} & 0 & 0 & 0\\
0 & (G_{0}^{N \Xi N})^{-1} & 0 & 0\\
0 & 0 & (G_{0}^{N \Lambda \Lambda})^{-1} & 0\\
0 & 0 & 0 & (G_{0}^{ \Lambda N \Lambda})^{-1}
\end{array}
\right) 
\\
\\
\left(
\begin{array}{cccc}
0 & (1-P_{23})P_{13}P_{23} & 0 & 0 \\
P_{12}P_{23} & -P_{13} & 0 & 0 \\
0 & 0 & 0 & (1-P_{23})P_{13}P_{23} \\
0 & 0 & P_{12}P_{23} & -P_{13}
\end{array}
\right)
\left(
\begin{array}{c}
\psi_{ \Xi NN} \\
\psi_{N \Xi N} \\
\psi_{N \Lambda \Lambda} \\
\psi_{ \Lambda N \Lambda}
\end{array}
\right) 
\\
\\
\equiv
\tilde{G_{0}^{-1}} \tilde{P} \tilde{ \psi}
\hfill
\end{array}
\label{5.8}
\end{equation}

\noindent
We recover the matrix $\tilde P$ of permutations already
occurring in Eq. (3.32) and in the coupled set (3.33).

Generally spoken that matrix Faddeev equation (3.33) is
a special case of Eq. (3.20) where  

\begin{equation}
\tilde{R} \equiv \tilde{P} \tilde{ \psi}
\label{5.10}
\end{equation}

\noindent 
contains now a general matrix $\tilde P$
of permutations (see below).
Now we can proceed and insert $\tilde \psi$ from (3.20)
 together with (5.9) back into (5.8) and
find

\begin{equation}
\tilde{U}=\tilde{G_{0}^{-1}} \tilde{P}( \tilde{ \psi^{0}}+ 
\tilde{G_{0}} \tilde{t} \tilde{P} \tilde{ \psi})=\tilde{G_{0}^{-1}} 
\tilde{P} \tilde{ \psi^{0}}+ \tilde{G_{0}^{-1}} \tilde{P} \tilde{G_{0}} 
\tilde{t} \tilde{G_{0}} \tilde{U}
\label{5.11}
\end{equation}

It remains to demonstrate that

\begin{equation}
\tilde{G_{0}^{-1}} \tilde{P}= \tilde{P} \tilde{G_{0}^{-1}}
\label{5.12}
\end{equation}

and thus 

\begin{equation}
\tilde{G_{0}^{-1}} \tilde{P} \tilde{G_{0}}= \tilde{P}
\label{5.13}
\end{equation}

Written out explicitely we have

\begin{equation}
\begin{array}{l}
\tilde{P} \tilde{G_{0}^{-1}}=  
\\
\\
{\scriptsize
\left(
\begin{array}{cccc}
0 & (1-P_{23})P_{13}P_{23}(E-H_{0}^{N \Xi N}) & 0 & 0 \\
P_{12}P_{23}(E-H_{0}^{ \Xi NN}) & -P_{13}(E-H_{0}^{N \Xi N}) & 0 & 0 \\
0 & 0 & 0 & (1-P_{23})P_{13}P_{23}(E-H_{0}^{ \Lambda N \Lambda}) \\
0 & 0 & P_{12}P_{23}(E-H_{0}^{N \Lambda \Lambda}) & -P_{13}
(E-H_{0}^{ \Lambda N \Lambda})
\end{array}
\right) }
\end{array}
\label{5.14}
\end{equation}

Let us regard two examples

\begin{equation}
\begin{array}{c}
(1-P_{23})P_{13}P_{23}H_{0}^{N \Xi N}(123)
\\
=(1-P_{23})H_{0}^{N \Xi N}(312)P_{13}P_{23} 
\\
=(1-P_{23})H_{0}^{ \Xi NN}(123)P_{13}P_{23}
\\
=H_{0}^{ \Xi NN}(123)(1-P_{23})P_{13}P_{23}
\end{array}
\label{5.15}
\end{equation}
\begin{equation}
P_{12}P_{23}H_{0}^{ \Xi NN}(123)=H_{0}^{ \Xi NN}(231)P_{12}P_{23}
=H_{0}^{N \Xi N}(123)P_{12}P_{23}
\label{5.16}
\end{equation}

\noindent
Corresponding relations are true in all cases and we find (5.12) to be 
valid. Thus we end up with the coupled matrix Faddeev equation

\begin{equation}
\tilde{U}= \tilde{P} \tilde{G_{0}^{-1}} \tilde{ \psi^{0}}+ \tilde{P} 
\tilde{t} \tilde{G_{0}} \tilde{U}
\label{5.17}
\end{equation}

For the case of the coupled $\Xi NN-\Lambda\Lambda N$ system they read
 in explicit form

\begin{equation}
U_{ \Xi d \Xi d}=(1-P_{23})P_{13}P_{23}(t_{ \Xi N ,\, \Xi N}G_{0}^{N \Xi N}
U_{N \Xi N , \Xi d}+t_{ \Xi N ,\, \Lambda \Lambda}G_{0}^{N \Lambda 
\Lambda}U_{N \Lambda \Lambda , \Xi d })
\label{5.18}
\end{equation}
\begin{equation}
\begin{array}{l}
U_{N \Xi N , \Xi d}=P_{12}P_{23}(G_{0}^{\Xi NN})^{-1} \phi_{ \Xi d}
\\
+P_{12}P_{23}t_{NN}G_{0}^{ \Xi NN}U_{ \Xi d \Xi d}-P_{13}
(t_{ \Xi N,\, \Xi N}G_{0}^{N \Xi N}U_{N \Xi N , \Xi d}
+t_{ \Xi N,\, \Lambda \Lambda}G_{0}^{N \Lambda \Lambda}
U_{N \Lambda \Lambda , \Xi d})
\end{array}
\label{5.19}
\end{equation}
\begin{equation}
U_{N \Lambda \Lambda , \Xi d}=(1-P_{23})P_{13}P_{23}
t_{N \Lambda ,\, N \Lambda}G_{0}^{ \Lambda N \Lambda}
U_{ \Lambda N \Lambda , \Xi d}
\label{5.20}
\end{equation}
\begin{equation}
U_{ \Lambda N \Lambda , \Xi d}=P_{12}P_{23}
(t_{ \Lambda \Lambda ,\, \Xi N}G_{0}^{N \Xi N}U_{N \Xi N , \Xi d}
+t_{ \Lambda \Lambda ,\, \Lambda \Lambda}G_{0}^{N \Lambda \Lambda}
U_{N \Lambda \Lambda , \Xi d})-P_{13}
t_{N \Lambda ,\, N \Lambda}G_{0}^{ \Lambda N \Lambda}
U_{ \Lambda N \Lambda , \Xi d}
\label{5.21}
\end{equation}

This is a generalized set of Alt-Grassberger-Sandhas (AGS) equations
 for that specific  system.
Obviously a corresponding  set can be written down for the coupled
 $\Lambda NN-\Sigma NN$ system based on the four Faddeev equations
 (3.25-3.27).

 The matrix equation (5.16) (generalized
AGS equations) is generally valid.
The proof is quite simple. In the set (3.19) the part including
 permutations
is by definition

\begin{equation}
<\alpha\beta\gamma |G_0 V P|\psi>=\sum_{\alpha '\beta '\gamma '}
\sum_{\alpha ''\beta ''\gamma ''}G_0^{\alpha\beta\gamma}
<\alpha\beta\gamma | V |\alpha '\beta '\gamma '>
<\alpha '\beta '\gamma ' | P |\alpha ''\beta ''\gamma ''>
\psi_{\alpha ''\beta ''\gamma ''}
\label{5.22}
\end{equation}

\noindent
 This defines the general  matrix for the permutation operators. Now
 $\tilde G_0\tilde P$
contains the matrix element

\begin{equation}
G_0^{\alpha\beta\gamma}
<\alpha\beta\gamma | P |\alpha '\beta '\gamma '>
=<\alpha\beta\gamma |G_0 P |\alpha '\beta '\gamma '>
\label{5.23}
\end{equation}

\noindent
The operator $G_0\equiv (E-H_0)^{-1}$
occurring in Eqs.~(3.4-3.5) is however fully symmetrical.
 Thus $G_0 P=PG_0$ and the relation (5.12) follows.

Finally we note that due to (5.8) and (4.4) the matrix $\tilde U$ of
 elastic
and  rearrangement quantities is related to the matrix $\tilde T$
 composed of partial
breakup amplitudes via

\begin{equation}
\tilde{U}= \tilde{P} \tilde{G_{0}^{-1}}\tilde{\psi^0}
+ \tilde{P} \tilde{T}
\label{5.24}
\end{equation}

\noindent
and can therefore be determined by quadrature once $\tilde T$
is evaluated from Eq. (4.5).
We remark that in the 3N system ( three identical particles) all those
matrix relations reduce to the well known single equations
 \cite{physrep}:

\begin{equation}
U=PG_{0}^{-1}\psi^0 +PT
\label{5.25}
\end{equation}
\begin{equation}
T=tP\psi^0 +tPG_{0}T
\label{5.26}
\end{equation}
\begin{equation}
U=PG_{0}^{-1}\psi^0 +P t G_0 U
\label{5.27}
\end{equation}

\noindent
In that form they have been solved rigorously (in a numerical sense)
for modern nuclear forces.

\section{Summary}

A symmetric Hamiltonian for distinguishable particles with different
 masses and non-symmetric interactions 
 has been constructed using 
the generalized Pauli principle based on particle labels
 added to the usual ones for space (momentum) 
coordinates and spin. This has been exemplified in two-body
 systems and then applied to three-baryon systems. Specifically
we had in mind the strangeness $S=-1$ and $-2$ systems.
We derived minimal sets of coupled Faddeev equations for bound states,
 for breakup
operators and rearrangement operators (generalized AGS equations).
In the case of the coupled $\Lambda NN-\Sigma NN$ system we arrive at
four Faddeev equations.
Dropping a possible $\Sigma$-admixture the system 
$\Xi NN-\Lambda\Lambda N$ is
also  described by four Faddeev
equations. Allowing for a $\Sigma$-admixture one would have nine equations.
An application of such a set of equations already appeared
in case of $^3_{\Lambda}$H in \cite{hypt}. An investigation of
 the  $S=-2$ system is planned.

\acknowledgements
We thank M.Oka for a stimulating discussion. This work was performed
during visits of W.G. at the Okayama University of Science and 
of K.M. at the Ruhr-University Bochum. Both authors thank for the 
hospitality extended to them at the two institutions. 

\newpage

\appendix

Appendix:  The property (4.16)

Starting from the Lippman-Schwinger equation

\begin{equation}
\begin{array}{c}
\left(
\begin{array}{cc}
t_{ \Xi N, \,\Xi N} & t_{ \Xi N, \,\Lambda \Lambda}  \\
t_{\Lambda \Lambda , \,\Xi N} 
& t_{ \Lambda \Lambda , \,\Lambda \Lambda} 
\end{array}
\right)
=
\left(
\begin{array}{cc}
V_{ \Xi N, \,\Xi N} & V_{ \Xi N, \,\Lambda \Lambda}  \\
V_{\Lambda \Lambda , \,\Xi N}^{tot} 
& V_{ \Lambda \Lambda , \,\Lambda \Lambda} 
\end{array}
\right)
\hfill\nonumber\\
+
\left(
\begin{array}{cc}
V_{ \Xi N, \,\Xi N} & V_{ \Xi N, \,\Lambda \Lambda}  \\
V_{\Lambda \Lambda , \,\Xi N}^{tot} 
& V_{ \Lambda \Lambda , \,\Lambda \Lambda} 
\end{array}
\right)
\left(
\begin{array}{cc}
G_0^{ \Xi NN} & 0  \\
0 & G_0^{\Lambda \Lambda N} 
\end{array}
\right)
\left(
\begin{array}{cc}
t_{ \Xi N, \,\Xi N} & t_{ \Xi N, \,\Lambda \Lambda}  \\
t_{\Lambda \Lambda , \,\Xi N} 
& t_{ \Lambda \Lambda , \,\Lambda \Lambda} 
\end{array}
\right)
\hfill
\end{array}
\end{equation}

\noindent 
one finds 

\begin{equation}
t_{\Lambda \Lambda , \,\Xi N}= V_{\Lambda \Lambda , \,\Xi N}^{tot}
+V_{\Lambda \Lambda , \,\Xi N}^{tot}\, G_0^{ \Xi NN}
\, t_{\Xi N , \,\Xi N}
+V_{\Lambda \Lambda , \,\Lambda \Lambda}\, G_0^{ \Lambda \Lambda N}
\, t_{\Lambda \Lambda , \,\Xi N}
\end{equation}

Here we number the interacting particles as 1 and 2.
Applying $P_{12}$ yields 

\begin{eqnarray}
P_{12}\, t_{\Lambda \Lambda , \,\Xi N}
&=&P_{12}\,  V_{\Lambda \Lambda , \,\Xi N}^{tot}
+
P_{12}\, V_{\Lambda \Lambda , \,\Xi N}^{tot}\, G_0^{ \Xi NN}
\, t_{\Xi N , \,\Xi N}
\nonumber
\\
&+&
P_{12}\, V_{\Lambda \Lambda , \,\Lambda \Lambda}\, G_0^{ \Lambda \Lambda N}
\, t_{\Lambda \Lambda , \,\Xi N}
\nonumber
\\
&=&- V_{\Lambda \Lambda , \,\Xi N}^{tot}
 - V_{\Lambda \Lambda , \,\Xi N}^{tot}\, G_0^{ \Xi NN}
    \, t_{\Xi N , \,\Xi N}
 + V_{\Lambda \Lambda , \,\Lambda \Lambda}\, G_0^{ \Lambda \Lambda N}
   P_{12}\, t_{\Lambda \Lambda , \,\Xi N}
\end{eqnarray}

We used the definition (2.16) and the obvious symmetry
of $V_{\Lambda\Lambda ,\Lambda\Lambda }G_0^{\Lambda\Lambda N}$
under exchange of particles 1 and 2.
It follows  

\begin{eqnarray}
P_{12}\, t_{\Lambda \Lambda , \,\Xi N}
&=&-(1-V_{\Lambda \Lambda , \,\Lambda \Lambda}\,
             G_0^{ \Lambda \Lambda N}\,)^{-1}
( V_{\Lambda \Lambda , \,\Xi N}^{tot}
 + V_{\Lambda \Lambda , \,\Xi N}^{tot}\, G_0^{ \Xi NN}
   \, t_{\Xi N , \,\Xi N} )
\nonumber
\\
&=&
-t_{\Lambda \Lambda , \,\Xi N}
\end{eqnarray}

which is claimed in (4.16).

\begin{figure}[htb]
   \vspace{2.9cm}
   \epsfysize=13cm
   \centerline{\epsffile{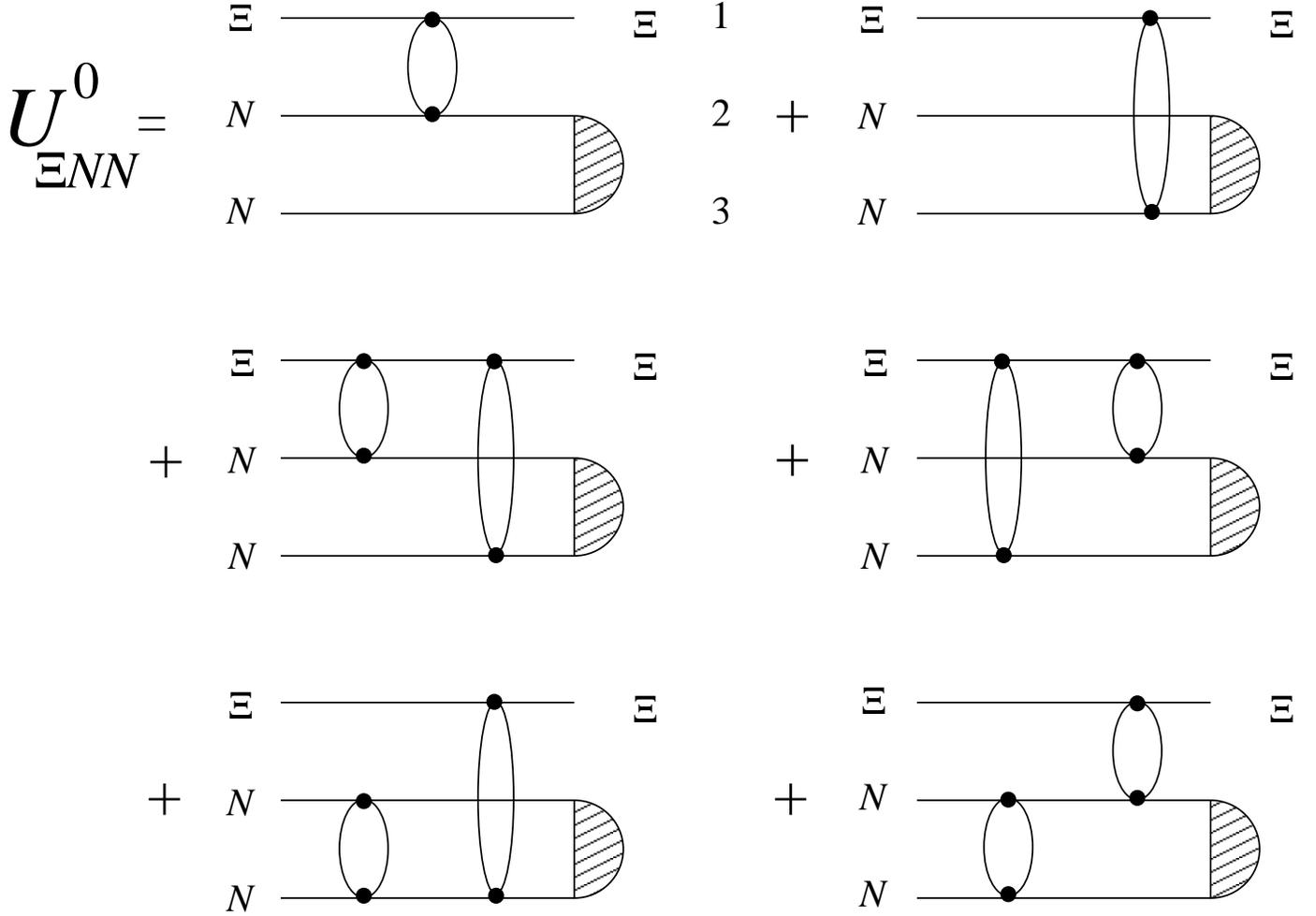}}
   \vspace{2cm}
   \centerline{\parbox{13cm}{\caption{\label{fig1a}
The $\Xi NN$  breakup process of Eq. (4.14) up to second
order in the $t$-matrices. The hatched half circle denotes
the incoming deuteron. 
  }}}
\end{figure}

\begin{figure}[htb]
   \vspace{2.9cm}
   \epsfysize=13cm
   \centerline{\epsffile{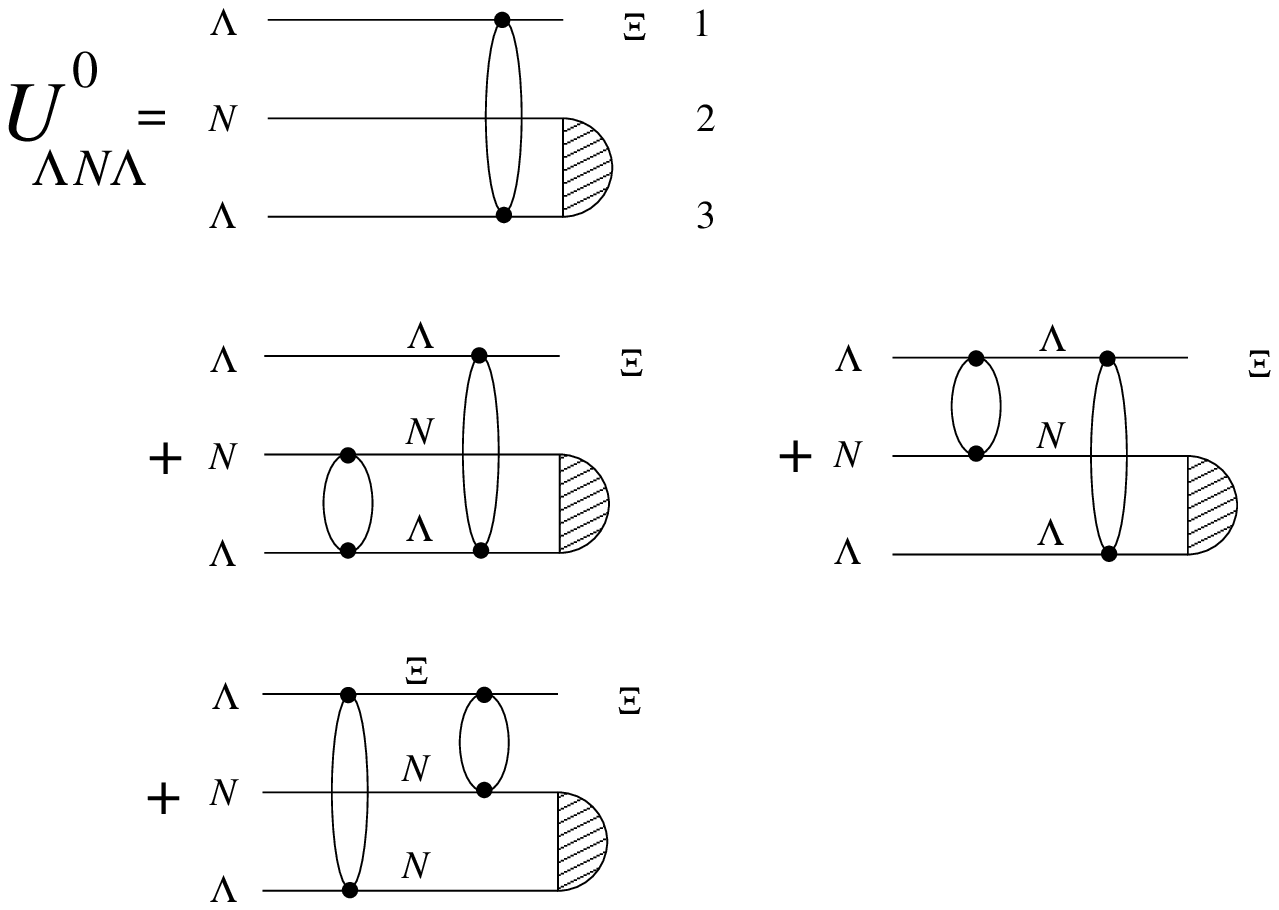}}
   \vspace{2cm}
   \centerline{\parbox{13cm}{\caption{\label{fig1b}
The $\Lambda N\Lambda$  breakup process of Eq. (4.15) up to
second order in the $t$-matrices. 
  }}}
\end{figure}


\begin{references}
\bibitem{ours}
K.~Miyagawa, H.~Kamada, and W.~Gl\"ockle,
Nucl. Phys. A614, 535 (1997).
\bibitem{afnan} S.B.Carr, I.R.Afnan, B.F.Gibson, 
Phys.Rev. C57(1998) 2858.
\bibitem{ags} E.O.Alt, P.Grassberger, W.Sandhas,
 Nucl. Phys. B2 (1967) 167
\bibitem{hypt} K.Miyagawa, H.Kamada, W.Gl\"ockle,
V.Stoks, Phys Rev. C51, 2905(1995).
\bibitem{hiyama} E.Hiyama, in Proceedings of the XVII
RCNP international symposium on innovative computational 
method in nuclear many-body problems, Osaka, November 1997,
eds. H.Horiuchi, M.Kamimura, H.Toki, Y.Fujiwara, M.Matsuo, 
Y.Sakuragi, World Scientific 1998, page 128. 
\bibitem{yamamu} H.Yamamura, K.Miyagawa, T.Mart, C.Bennhold,
H.Haberzettl, W.Gl\"ockle, Phys. Rev. C61, 014001-1(1999).
\bibitem{reinhold} D.M.Koltenuk, PhD thesis, University
 of Pennsylvania, 1999, unpublished.
\bibitem{lee} T.-S.H. Lee, V.Stoks, B.Saghai, C.Fayard, 
Nucl. Phys. A639, 247c (1998).
\bibitem{glockle} W.Gl\"ockle, The Quantum Mechanical Few-Body
Problem, Springer Verlag, 1983.
\bibitem{polonica} D.H\"uber, H.Kamada, H.Witala, W.Gl\"ockle,
Acta Phys. Polonica B21, 1677(1997).
\bibitem{physrep} W.Gl\"ockle, H.Witala, D.H\"uber, H.Kamada, J.Golak,
Phys. Rep. 274, 107(1996).
\bibitem{zeit} W.Gl\"ockle, Zeitschrift f.Phys. 271, 31(1974).
\end{references}
\end{document}